\documentclass[paper]{JINST}
\usepackage{amsmath,dcolumn}

\title{Low-energy negative muon interaction with matter}

 \author{
 Petar Danev$^a$\thanks{Corresponding author.},
 Andrzej Adamczak$^b$,
 Dimitar Bakalov$^a$,
 Emiliano Mocchiutti$^c$,
 Mihail Stoilov$^a$,
 and Andrea Vacchi$^{d,c}$\\
 \llap{$^a$}Institute for Nuclear Research and Nuclear Energy, Bulgarian Academy of Sciences,\\
  blvd. Tsarigradsko ch. 72, Sofia 1142, Bulgaria\\
  E-mail: \email{petar\_danev@abv.bg}\\
 \llap{$^b$}Institute of Nuclear Physics, Polish Academy of Sciences,\\
  Radzikowskiego 152, PL31342 Krak\'{o}w, Poland\\
  \llap{$^c$}Istituto Nazionale di Fisica Nucleare,\\
  Padriciano 99, AREA Science Park, 34149 Trieste, Italy\\
  \llap{$^d$}Mathematics and Informatics Department,
  Udine University\\ via delle Scienze 206,Udine\\
  }

\abstract{
 Using simulated data, obtained with the {\sc fluka} code, we derive
 empirical regularities  about the propagation and stopping of
 low-energy negative muons in hydrogen and selected solid
 materials.
 The results are intended to help the
 preliminary stages of the set-up design for
 experimental studies of muon capture and muonic atom spectroscopy.
 Provided are approximate expressions for the parameters of the
 the momentum, spatial and angular distribution of the
 propagating muons.
 In comparison with the available data on the stopping power and
 range of muons (with which they agree in the considered energy range)
 these results have the advantage to also describe the statistical
 spread of the muon characteristics of interest.
}


\begin{document}

\section{Introduction}

 The laser spectroscopy measurement of the hyperfine splitting in the ground state of
 muonic hydrogen, considered as fundamental test of QED, complementary to the
 measurement in ordinary hydrogen \cite{essen}, has been a major
 experimental challenge for more than two decades \cite{pla92,hfi01}.
 The interest in it grew up significantly when the muonic hydrogen Lamb
 shift experiment revealed a 9$\sigma$ discrepancy
 between the proton charge radius values extracted from muonic
 hydrogen spectroscopy and $e-p$ scattering data \cite{pohl}. The point
 was that from the hyperfine splitting of muonic hydrogen one can directly
 extract the value of the Zemach radius of the proton \cite{zemach},
 juxtapose it to the value extracted from ordinary hydrogen
 spectroscopy \cite{pra03} and this way test most of the hypotheses put forward
 to explain the proton size puzzle. The FAMU collaboration
 \cite{famu} is currently preparing an experiment that uses a method based on the
 study of the diffusion of the hydrogen muonic atoms in
 appropriate gas target and its response to monochromatic laser
 radiation of resonance frequency \cite{nimb,pla15}. An alternative
 experimental approach has been recently suggested in \cite{ishida1}.

 As discussed in details in \cite{hfi01,nimb,pla15}, the
 efficiency of this method is determined
 by the energy dependence in the epithermal range of the rate of muon transfer
 in collisions of the muonic hydrogen atoms with the atoms of
 the heavier gas admixture. There are experimental indications
 that muon transfer to oxygen has the needed
 characteristics \cite{werth}, in agreement with the theoretical
 estimates \cite{dupays,cdlin}, but the experimental accuracy is
 far from being sufficient for planning and
 optimizing the measurement of the hyperfine splitting. Because of
 this, an experiment uniquely dedicated to the thorough investigation of
 the collision energy dependence of the rate of muon transfer to
 various gases was launched as a first stage of the FAMU project.
 The measurements are to be performed at the pulsed muon source of
 the RAL-RIKEN facility \cite{RAL}. Muons of initial momentum in
 the 50-70 MeV/c range will be stopped in a mixture of hydrogen
 and various heavier gases at high pressure, and the time distribution of the
 characteristic X-rays signalling the transfer of the muon to the
 admixture nuclei will be registered and analyzed using the algorithms of
 \cite{pla15}.

 One of the main challenges in the preparation of the set-up
 for the above experiment, as well as in other experiments
 studying muon capture by protons or muonic atom spectroscopy,
 is the design of the gas target which has to
 satisfy the following conditions:
 \begin{enumerate}
   \item as much as possible of the incident muons are stopped in gas, i.e.
   the losses in the front, side and rear walls of the gas
   container are minimized;

   \item as large part as possible of the emitted characteristic X-rays
   reach the radiation detectors around the gas target that, in principle,
   cover only a small fraction of the solid angle\label{compact};
%
%   \item the background, mainly due to the electrons produced in
%   muon decay, is suppressed as strongly as possible.
 \end{enumerate}

 Cond.~1 requires the detailed study of the balance between the
 stopping power of the target as function of the muon momentum
 and the pressure (with account of the dependence of the
 muon flux on the initial momentum), and the
 losses in the walls that depend on their
 composition and thickness. Cond.~2 requires the study
 of the spatial aspects of the muon stopping and the formation
 of muonic atoms.
% Finally, Cond.~3 needs the interaction of the muons with the
% container wall ingredients to be taken into account.
 %
 During the search for the optimal target and detector set-up
 a large variety of geometrical configurations and materials were
 considered and investigated with Monte Carlo simulation codes.
 We noticed that blind iterative simulations are not necessarily
 the best approach and that it is useful and illuminating to
 have an analytic formulation allowing to identify the optimal
 path before proceeding with cross-over simulations.
 Leaving the detailed description of the selected experimental
 design for the muon transfer experiment to be reported elsewhere,
 we present here some
 characteristics and regularities of the propagation of negative
 muons in materials, established empirically with the {\sc fluka} simulation code
 \cite{fluka1,fluka2}, which we found particularly useful in the
 preliminary stages of the set-up design and, we believe, are
 applicable to a broad range of muon physics modelling problems.

 We focus our attention on two ``elemental'' cases: (a) a
 monochromatic collinear muon beam normally incident on a solid
 homogeneous layer made out of some of the materials of interest
 (steel, aluminum, gold, and polystyrene), and (b) a monochromatic collinear muon
 beam stopped in an unbound domain filled with hydrogen.
 In case (a), only part $Q, Q\le1$ of the incident muons cross the
 layer; they are scattered at angle $\Theta$ with
 final momentum $p'$, in general different from the initial momentum $p$. We
 studied the probability distributions of $p'$ and $\Theta$ as
 functions of $p$ and the layer thickness $d$ and derived simple
 approximate expressions for the mean and root mean squared
 (r.m.s.) deviation values, as well for the surviving rate $Q$.
 These expressions
 were verified to provide satisfactory accuracy for $p\le75$
 MeV/c - the range of interest of initial momenta available at the RAL-RIKEN
 facility.
 In case (b) the muons are slowed in collisions with the hydrogen
 molecules and then stopped and captured in muonic hydrogen
 atoms. We investigated the spatial distribution of the stopping
 points and derived simple
 approximate expressions for the mean value and root mean squared
 deviation of the cylindrical coordinates $z$ and $r$ of the
 stopping points as functions of the initial momentum $p$ and the
 hydrogen gas pressure $H$.
 In principle, these formulae could be used to model the propagation
 and stopping of non-monochromatic muon beams in complex
 geometrical configurations, but undoubtedly such an approach will
 be much less efficient and accurate that the direct Monte Carlo
 simulations.
 The reported results were intended only -- and shown to be -- a convenient tool for the
 preliminary estimation of the impact of individual elements during
 the process of designing complex set-ups for the experimental
 study of muonic atoms.

 \section{Propagation of negative muons across solid material layers}
 \label{sec-layer}

\begin{figure}[ht]
 \begin{center}
 \includegraphics[width=.8\textwidth]{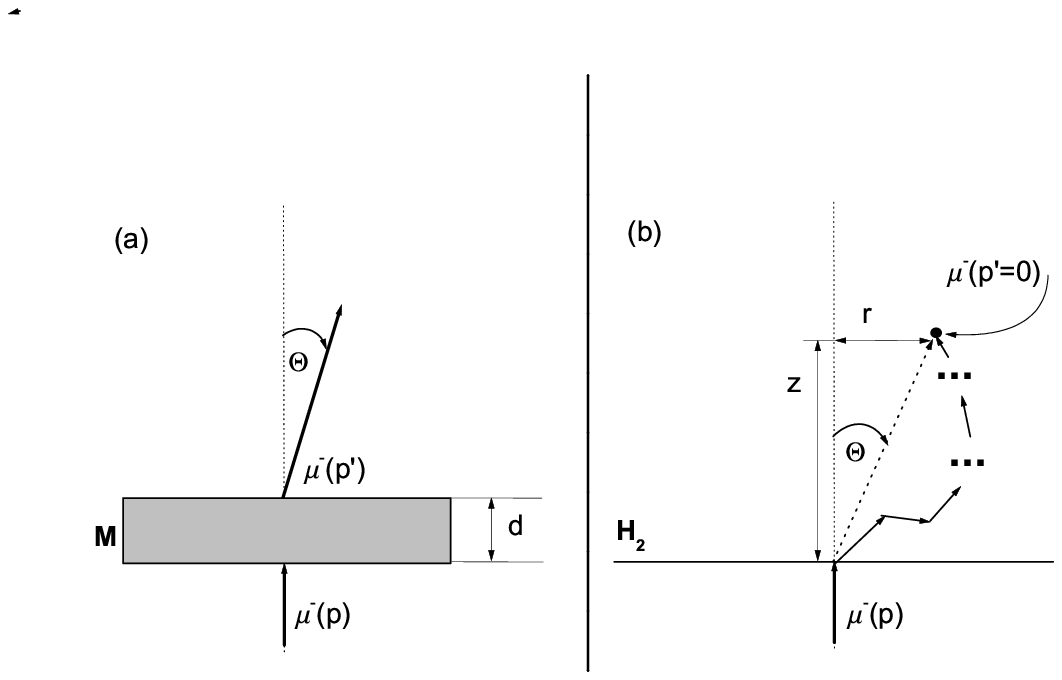}
 \caption{Geometry of the simulated events.
 (a) Muons with momentum $p$ cross normally the boundary of a layer
 of material {\bf M} with thickness $d$.
 The momentum $p'$ and the scattering angle $\Theta$ at the
 crossing with the opposite boundary are evaluated using the  {\sc fluka}
 code. (b) Muons with initial momentum $p$ propagate in hydrogen
 medium at gas pressure $H$. The coordinates $z$, $r$, and $\Theta$
 of the muon stopping points are evaluated with  {\sc fluka}.}
 \label{geom}
 \end{center}
 \end{figure}

 In this section we consider the interaction of negative muons with
 four materials of interest: Aluminum (denoted by {\bf A}),
 stainless steel 316LN \cite{SS316LN} ({\bf S}),
 Gold ({\bf G}), and Polystyrene ({\bf P}) using
 simulated results obtained with the  {\sc fluka} code
 \cite{fluka1,fluka2}.
 In each run monochromatic bunches of $N=10^5$ muons with
 initial momentum $p, 15\le p\le75$ MeV/c are launched against a layer of
 material {\bf M}={\bf A}, {\bf S}, {\bf G} or {\bf P} with thickness $d$
 along the $z$-axis, normal to the layer
 surface (see Figure~\ref{geom}a).

 \subsection{Muon survival probability}
 \label{p0}

\begin{figure}[ht]
 \begin{center}
 \includegraphics[width=.8\textwidth]{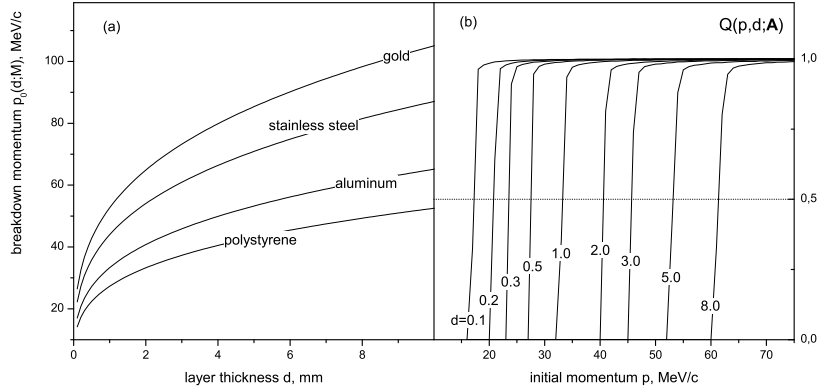}
 \caption{Interaction of a muon beam with solid material layers.
 (a) The breakdown momentum $p_0(d;{\bf M})$ versus the thickness
 $d$ of the layer of material {\bf M}. (b) Probability $Q(p,d;{\bf
 A})$ that a normally incident muon with initial momentum $p$ is
 not stopped in aluminum layer of thickness $d, 0.1\le d\le8$ mm.
}
 \label{stip2}
 \end{center}
 \end{figure}

 Denote the number of muons, stopped within the layer, by $N_{0}(p,d;{\mathbf M})$,
 $0\le N_{0}(p,d;{\mathbf M})\le N$.
 In terms of the latter, the empirical
 probability distribution
 for a muon with initial momentum $p$ to cross a layer of thickness $d$ is
 $Q(p,d;{\mathbf M})=1-N_{0}(p,d;{\mathbf M})/N$.
 Figure~\ref{stip2} shows ``the rate of survival'' $Q(p,d;{\bf A})$ of a
 monochromatic muon beam versus the initial momentum $p$, for a set of
 values of the Aluminum layer thickness $d$.
 For momenta below
 the breakdown thickness-dependent value $p_0(d;{\mathbf M})$ practically all the muons are
 stopped in the Al layer; for momenta above $p_0(d;{\mathbf M})$ practically
 all muons pass through the layer. The interval of $p$ within which $Q$
 rises from $Q\sim20$\% to $Q\sim$80\% is as narrow as 1 MeV/c.
 We defined the breakdown momentum
 by $Q(p_0(d;{\mathbf M}),d;{\mathbf M})=0.5$, evaluated
 $p_0(d;{\bf M})$ for a set of values of the thickness $d$ between 0.01 mm and
 8 mm and found out that the following 2-parameter
 expression fits very well the calculated values:
 \begin{equation}
   p_0(d;{\bf M})=a_1 d^{a_2}
   \label{fitsteep}
 \end{equation}
 The values of the parameters $a_1$ and $a_2$ for the
 material of interest are given in
 Table~\ref{tab-steep}.

 \begin{table}[h]
 \begin{center}
 \caption{Numerical values of the coefficients $a_i,i=1,2$
 in the fitting expression for the breakdown momenta $p_0$  (\ref{fitsteep}). }
 \begin{tabular}{c|rrrr}
 \hline
% material & polystyrene & aluminum & steel S316LN & gold \\ \hline
 {\bf M} & {\bf P} & {\bf A} & {\bf S} & {\bf G} \\ \hline
 $a_1$, MeV/c & 27.3 & 33.3 & 44.0 & 52.7
 \\
 $a_2$\phantom{, MeV/c} & 0.2842 & 0.2916 & 0.2964 & 0.2997
 \\
 $\rho$, g cm$^{-3}$ & 1.03 & 2.7 & 7.99 & 19.29
 \end{tabular}
 \label{tab-steep}
 \end{center}
 \end{table}
 It is worth mentioning that the values of the breakdown momentum
 %, calculated with  {\sc fluka}
 $p_0$ for the four materials listed above
 are quite accurately fitted with the single 3-parameter expression
 \begin{equation}
   p_0(d,{\bf M})=a_1 d^{a_2} \rho_{\bf M}^{a_3},
   \label{fit-step2}
 \end{equation}
 where $\rho_{\bf M}$ is the density of the material ${\bf M}$ (in
 g cm$^{-3}$, see Table~\ref{tab-steep}), $d$ is the thickness (in mm), and $a_1=26.6$ MeV/c,
 $a_2=0.2969$, and $a_3=0.2342$.
 %Note, however, that Eq.~(\ref{fit-step2}) has not been verified for other materials.

 \subsection{Momentum and angular composition of the scattered muon beam}

\begin{figure}[h]
 \begin{center}
 \includegraphics[width=.8\textwidth,height=.4\textheight]{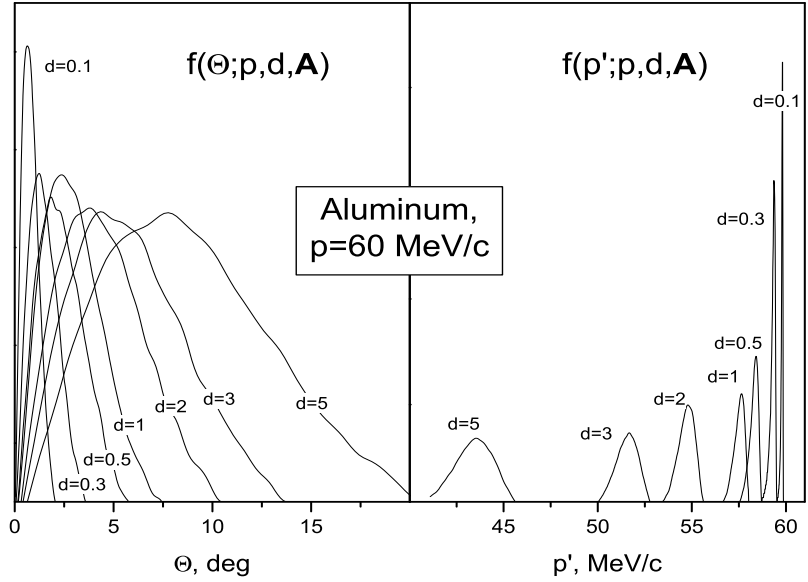}
 \caption{
 (a) Probability density of the distribution of the scattering
 angle $\Theta$ of the outgoing muons with initial momentum $p=60$ MeV/c,
 scattered by an aluminum layer of thickness $d$,
 in arbitrary units; (b) Same for the probability density of the distribution of the
 final momentum $p'$.}
 \label{plotbin60}
 \end{center}
 \end{figure}

\begin{figure}[h]
 \begin{center}
 \includegraphics[width=.8\textwidth]{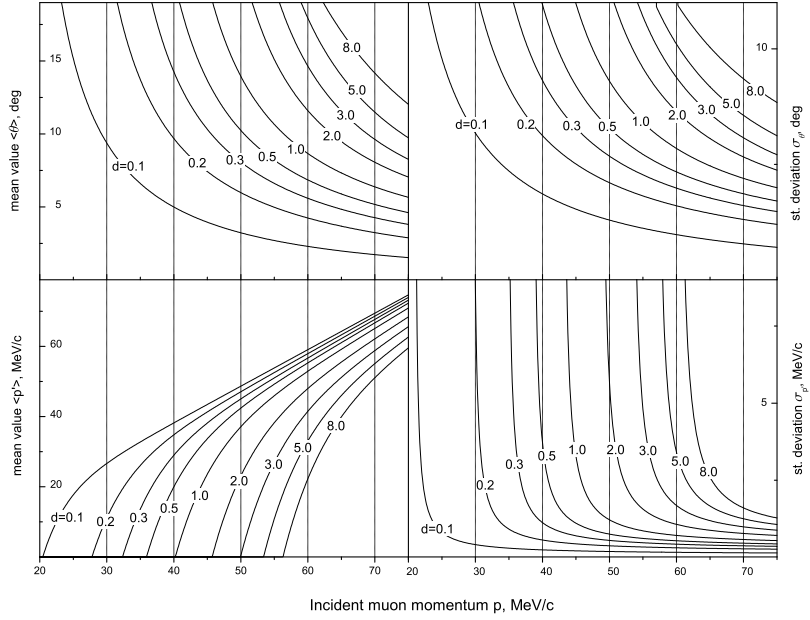}
 \caption{Momentum spectrum and angular profile of outgoing
 muon beams. Plotted are
 the mean values $\langle p'\rangle$, $\langle\Theta\rangle$
 and the r.m.s. deviations $\sigma_{p'}$, $\sigma_{\Theta}$ of $p'$
 and $\Theta$ vs. the initial momentum $p$, for aluminum layer of
 thickness $d, 0.1\le d\le8$ mm.}
 \label{al4plots}
 \end{center}
 \end{figure}

 From the simulated data on the final momentum $p'$ and the scattering
 angle $\Theta$, obtained with  {\sc fluka} for the $N_1=N-N_0$ muons that
 cross the layer (see Figure~\ref{geom}) we evaluated the empirical
 probability densities $f(p';d,p,{\bf M})$ and $f(\Theta;d,p,{\bf
 M})$ of the final momentum and angular distribution of the
 outgoing muons, as well as the mean values $\langle p'\rangle$,
 $\langle\Theta\rangle$ and the root mean square
 deviations $\sigma_{p'}$, $\sigma_{\Theta}$ of the final momentum $p'$ and
 the scattering angle $\Theta$
 as functions of the initial momentum $p$, the
 thickness $d$ and the material {\bf M} of the layer.
 Figure~\ref{plotbin60} illustrates the shape of the probability density of
 the momentum and angular distributions of an incident muon beam
 with $p=$60 MeV/c for aluminum layer thickness in the range
 $0.1\le d\le 8$ mm. Figure~\ref{al4plots} shows the dependence of
 the mean and root mean square values for these distributions on
 the initial muon momentum $p$.
 %--------
  %---------------
 %----------------------
 To make these results helpful for practical use,
 we fitted the calculated values of
 $\langle p'\rangle$, $\langle\Theta\rangle$,
 $\sigma_{p'}$, and $\sigma_{\Theta}$
 for incident muon momentum
 $15\le p\le75$ MeV/c with the following approximants:
 \begin{eqnarray}
   &\hspace*{-1cm}
   \langle p'\rangle(p,d;{\bf M})=&
   c^{(1)}_1 d(1-d^2)+p\,(c^{(1)}_2 d^2+c^{(1)}_3 d+1)+
   \label{pav}\\
   &&
   (c^{(1)}_4 d^2+c^{(1)}_5 d+c^{(1)}_6)
%   d^6\exp(-c^{(1)}_7(p-48))
   \exp(-c^{(1)}_7(p-48-6\log d))
   \nonumber
   \\
   &\hspace*{-1cm}
   \sigma_{p'}(p,d;{\bf M})=&
%   c^{(2)}_1 \sqrt{d}+
%   (c^{(2)}_2+c^{(2)}_3 \sqrt{d})/(p-c^{(2)}_4-c^{(2)}_5 d^{c^{(2)}_6})
   c^{(2)}_1+c^{(2)}_2\!\sqrt{d}+(c^{(2)}_3+c^{(2)}_4\!\sqrt{d}+c^{(2)}_5\!\sqrt{p})/
   \sqrt{c^{(2)}_6+(c^{(2)}_7+c^{(2)}_8 d^{0.3}+p)^2}
%   \left(c^{(2)}_6+(c^{(2)}_7+c^{(2)}_8 d^{0.3}+p)^2\right)^{-1/2}
   \label{sigmap}
   \\
   &\hspace*{-1cm}
   \langle \Theta\rangle(p,d;{\bf M})=&
   \bigg(
   \underbrace{\sum\limits_{i=0}^3\sum\limits_{j=0}^2}_{i+j\le3}\,c^{(3)}_{ij}
   d^i\,p^j+
   \bar{c}^{(3)}_1 p^2/d^2+
   \bar{c}^{(3)}_2\sqrt{d} p^2+
   \bar{c}^{(3)}_3 d^3/\sqrt{p}\bigg)^{-1}
   \label{thav}
   \\
   &\hspace*{-1cm}
   \sigma_{\Theta}(p,d;{\bf M})=&
   \bigg(
   \underbrace{\sum\limits_{i=0}^3\sum\limits_{j=0}^2}_{i+j\le3}\,c^{(4)}_{ij}
   d^i\,p^j+
   \bar{c}^{(4)}_1 p^2/d^2+
   \bar{c}^{(4)}_2\sqrt{d} p^2+
   \bar{c}^{(4)}_3 d^3/\sqrt{p}\bigg)^{-1}
   \label{sigmath}
  \end{eqnarray}
 The numerical values of the parameters in these fitting formulae
 for the materials of interest {\bf M}={\bf P} (polystyrene),
 {\bf A} (aluminum), {\bf S} (stainless steel SS316LN), and {\bf G} (gold), are
 given in Tables \ref{fit-meanp}, \ref{fit-sigmap}, and
 \ref{fitz}. The quality of the fit is described with the value of the mean squared deviation
 \begin{equation}
   \delta=\left(K^{-1}\sum\limits_{k=1}^K
   (V_k/F(p_k,d_k)-1)^2\right)^{1/2},
   \label{allk}
 \end{equation}
 where the summation is over all $K\sim400$ pairs of values of the parameters
 ($p$ and $d$ in this case) for which the values $V_k$ have been calculated
 with the  {\sc fluka} code; $F(p_k,d_k)$ denotes the value of the
 fitting function.
 Each run of {\sc fluka} used a sample of $10^5$ muons,
 so that the statistical uncertainty of $V_k$
 does not exceed 0.4\% and can be neglected
 with respect to $\delta$.
 The numerical uncertainty of the fitting expression coefficients is
 below $10^{-4}$, but typing them with more than 4
 digits would be in excess of the overall precision.

 \section{Stopping negative muons in gaseous hydrogen}
 \label{sec-gas}

\begin{figure}[ht]
 \begin{center}
 \includegraphics[width=.8\textwidth]{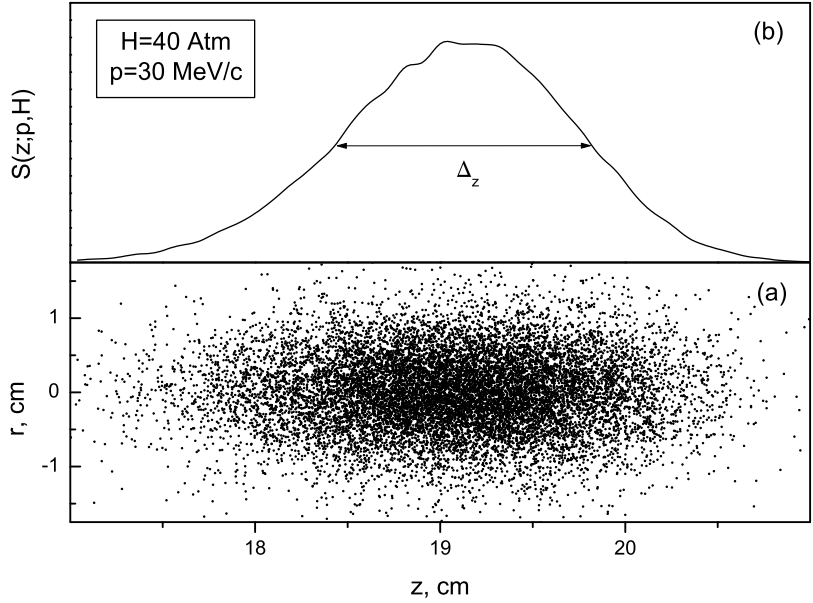}
 \caption{(a) Spatial distribution of the muon stopping positions for a
 collinear beam with initial momentum $p=$30 MeV/c in pure
 hydrogen at pressure 40 Atm and temperature 300K.
 (b) Longitudinal density $S(z;p,H)$, in arbitrary units.
 $\Delta_z$ denotes the FWHH size of the stopping area.}
 \label{scat}
 \end{center}
 \end{figure}

\begin{figure}[ht]
 \begin{center}
 \includegraphics[width=.8\textwidth]{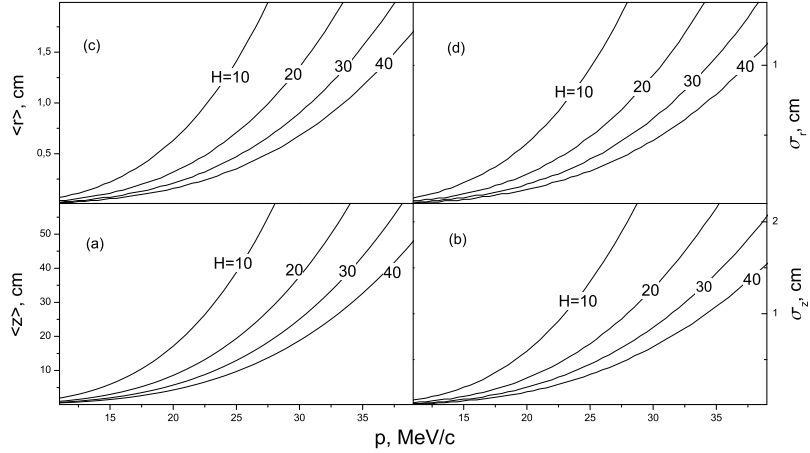}
 \caption{Muon stopping in pure hydrogen.
 Dependence of the mean
 longitudinal and transversal $\langle z\rangle(p,H)$, $\langle r\rangle(p,H)$
 and their r.m.s. deviations $\sigma_{z}(p,H)$, $\sigma_{r}(p,H)$ vs.
 the initial momentum $p$, for hydrogen pressure $H=10(10)40$ Atm.}
 \label{fig-hyd4plots}
 \end{center}
 \end{figure}

 In this section we use the  {\sc fluka} code to simulate the propagation of negative muons in
 hydrogen gas target. In each run monochromatic collinear bunches
 of $N=10^5$ muons with initial momentum $p, 1\le p\le 60$ MeV/c are
 launched in the gas with pressure $H, 5\le H\le 40$ Atm, the
 cylindrical coordinates $z_i$ and $r_i$  of the end points
 $T_i,i=1,\ldots,N$ of the muon trajectories
 (where the muons are stopped and supposedly immediately captured in a muonic
 hydrogen atom, see Figure \ref{geom}) are registered, and on this basis the empirical
 density of the spatial distribution of the muon stop points,
 $s(z,r;p,H)$ is evaluated as function of the initial momentum $p$
 and the hydrogen gas pressure $H$ (assuming axial symmetry).
 Figure~\ref{scat}a is the scatter plot
 of the set of stopping points for a muon bunch with initial
 momentum $p=30$ MeV/c, propagating in pure hydrogen at $H=40$
 Atm along the $z$-axis. Most of the muons are stopped at about 19 cm
 from the entry point, with a spread of about $\pm1$ cm in both
 longitudinal and transversal directions. Figure~\ref{scat}b
 presents the longitudinal density
 $S(z;p,H)=\int s(z,r;p,H)\,r\,dr$.
 As in Section \ref{sec-layer}, we evaluated
 the mean values and the r.m.s. deviations
 of $z$ and $r$ as functions of the initial momentum $p$ and the
 hydrogen pressure $H$:
 \begin{eqnarray}
   &&\langle z\rangle(p,H)=\iint z\,\,s(z,r;p,H)\,dz\,\,r\,dr
   \\
   &&
   %\ \
   \sigma_z(p,H)=\big(\iint \left(z-\langle z\rangle(p,H)\right)^2
   s(z,r;p,H)\, dz\,\,r\,dr\big)^{1/2}
   %\\
   %&&\langle r\rangle(p,H)=\int r s(z,r;p,H) dz\,r\,dr\\
   %&&\sigma_r(p,H)=\big(\int (r-\langle r\rangle(p,H))^2 s(z,r;p,H) dz\,r\,dr\big)^{1/2}
 \end{eqnarray}
 (and similar for $\langle r\rangle(p,H)$ and
 $\sigma_r(p,H)$).
 Figure \ref{fig-hyd4plots} illustrates the dependence of these
 quantities on the initial muon momentum $p$ for hydrogen
 pressures $H=10(10)40$ Atm.
 The functional dependence is fitted with expressions of the form:
 \begin{equation}
   b^{(k)}_1/H\,p^{b^{(k)}_2}\,\left(1+b^{(k)}_3\,p^2\exp(-b^{(k)}_4
   p)\right), k=1,\ldots,4,
   \label{eq-fithyd}
 \end{equation}
 where the index $k=1,2,3,4$ labels the coefficients of the fit of
 $\langle z\rangle(p,H)$, $\sigma_z(p,H)$, $\langle
 r\rangle(p,H)$, and $\sigma_r(p,H)$, respectively. The numerical
 values of $b^{(k)}_i$ are given in Table ~\ref{tab-fithyd}.

 \begin{table}[p]
 \begin{center}
 \begin{footnotesize}
 \caption{Coefficients $c^{(1)_i},i=1,\ldots,7$ in the fit of
 Eq.~(\ref{pav}).}
 \label{fit-meanp}
 \begin{tabular}{c|D{.}{.}{5}D{.}{.}{4}D{.}{.}{4}D{.}{.}{4}D{.}{.}{4}D{.}{.}{4}
 D{.}{.}{4}D{.}{.}{6}D{.}{.}{1}@{\hspace{3mm}}|}
% \begin{tabular}{c|rr|rr|rr|rr|}
 {\bf M} &
 \multicolumn{1}{c}{$c^{(1)}_1$} &
 \multicolumn{1}{c}{$c^{(1)}_2$} &
 \multicolumn{1}{c}{$c^{(1)}_3$} &
 \multicolumn{1}{c}{$c^{(1)}_4$} &
 \multicolumn{1}{c}{$c^{(1)}_5$} &
 \multicolumn{1}{c}{$c^{(1)}_6$} &
 \multicolumn{1}{c}{$c^{(1)}_7$} &
% \multicolumn{1}{c}{$c^{(1)}_8$} &
 \multicolumn{1}{l}{$\delta$, \%}
  \\
 \hline

 {\bf P} &
 0.008314 & 0.002309 & -0.02480 & -0.02269 & -0.3106 & -0.2644 &
 0.1497 & 2.0 \\
 {\bf A} &
 0.02184 & 0.004111 & -0.03340 & -0.2112 & -1.597 & -0.6895 &
 0.1245 & 2.0 \\
 {\bf S} &
 0.4158 & 0.03141 & -0.09804 & -3.010 & -8.087 & -1.417 & 0.1205 &
 1.1 \\
 {\bf G} &
 1.802 & 0.06397 & -0.1427 & -5.318 & -27.07 & -1.597 & 0.09386 &
 0.7
 %\\
 %
 \end{tabular}
 \end{footnotesize}
 \end{center}
 \end{table}

 \begin{table}[p]
 \begin{center}
 \begin{footnotesize}
 \caption{Coefficients $c^{(2)_i},i=1,\ldots,8$ in the fit  of
 Eq.~(\ref{sigmap}).}
 \label{fit-sigmap}
 \begin{tabular}{c|D{.}{.}{4}D{.}{.}{4}D{.}{.}{4}D{.}{.}{4}D{.}{.}{5}
 D{.}{.}{2}D{.}{.}{4}D{.}{.}{2}D{.}{.}{2}}%@{\hspace{1mm}}}
 {\bf M} &
 \multicolumn{1}{c}{$c^{(2)}_1$} &
 \multicolumn{1}{c}{$c^{(2)}_2$} &
 \multicolumn{1}{c}{$c^{(2)}_3$} &
 \multicolumn{1}{c}{$c^{(2)}_4$} &
 \multicolumn{1}{c}{$c^{(2)}_5$} &
 \multicolumn{1}{c}{$c^{(2)}_6$} &
 \multicolumn{1}{c}{$c^{(2)}_7$} &
 \multicolumn{1}{c}{$c^{(2)}_8$} &
 \multicolumn{1}{r}{$\delta$, \%}
  \\
 \hline
 {\bf P} & -0.03 & 0.1614 & -1.184 & 0.9827 & 0.4001 & 0.1 &
 -1.346 & -25.58 & 4.3 \\
 %0.1523 & 0.6329 & 1.470 & -2.521  & 29.25 & 0.27 & 3.9 \\
 %
 {\bf A} & -0.01 & 0.2456 & 1.001 & 2.279 & -0.04457 & 0.1 &
 -0.1957 & -32.47 & 4.7 \\
 %
 %0.2317 & 0.6877 & 2.370 &  0.2504 & 32.37 & 0.30 & 4.9 \\
 %
 {\bf S} & 0.01 & 0.4014 & 3.568 & 5.566 & -0.5351 & 1.0 & 0.5765
 & -43.90 & 3.1 \\
 %0.3450 & 1.094  & 5.125 & -1.003  & 43.90 & 0.30 & 3.1 \\
 %
 {\bf G} & 0.047 & 0.3722 & 3.796 & 18.64 & -0.7801 & 0.25 & 2.116
 & -53.54 & 4.5
 %0.3478 & 0.3173 & 16.88 &  1.538  & 49.76 & 0.33 & 3.8
 \end{tabular}
 \end{footnotesize}
 \end{center}
 \end{table}

 \begin{table}[p]
 \begin{center}
 \begin{footnotesize}
 \caption{Coefficients $c^{(k)}_{ij}$ and $\bar{c}^{(k)}_{i}, i=1,2,3$,\ $k=3,4$ in the fit
 of Eqs.~(\ref{thav},\ref{sigmath}). $a[b]$ stands for $a.10^b$.}
 \label{fitz}
 \begin{tabular}{l|D{.}{.}{4}D{.}{.}{4}D{.}{.}{4}D{.}{.}{4}D{.}{.}{4}D{.}{.}{4}
 D{.}{.}{4}D{.}{.}{4}@{\hspace{5mm}}|}
% \begin{tabular}{c|rr|rr|rr|rr|}
 \hline
% {\bf A} & -3.554 & -17.35  & 8.686 & 1.800 & 0.4831 & 0.2674 & -0.1986 &
%+0.01372 & 0.0000317 & -0.02489 & 0.01136 & -12.86
 & \multicolumn{2}{c}{{\bf P} (Polystyrene)}
 & \multicolumn{2}{c}{{\bf A} (Aluminum)}
 & \multicolumn{2}{c}{{\bf S} (Stainless steel)}
 & \multicolumn{2}{c}{{\bf G} (Gold)}
 \\
 & \multicolumn{1}{c}{$\langle\Theta\rangle,\,k=3$}
 & \multicolumn{1}{c}{$\sigma_{\Theta},\,k=4$}
 & \multicolumn{1}{c}{$\langle\Theta\rangle,\,k=3$}
 & \multicolumn{1}{c}{$\sigma_{\Theta},\,k=4$}
 & \multicolumn{1}{c}{$\langle\Theta\rangle,\,k=3$}
 & \multicolumn{1}{c}{$\sigma_{\Theta},\,k=4$}
 & \multicolumn{1}{c}{$\langle\Theta\rangle,\,k=3$}
 & \multicolumn{1}{c}{$\sigma_{\Theta},\,k=4$}
 \\
 \hline
  {\vrule height 4mm width 0cm}
  $c^{(k)}_{00}$ & -8.338 & -4.064 & -5.192 & -3.554        & -3.687    & -6.709    & -1.672    &
    -1.831    \\
  {\vrule height 5mm width 0cm}
  $c^{(k)}_{10}$ & -13.14 & -16.92 & -11.68 & -17.35        & -11.79    & -20.63    & -33.77   &
    -52.53    \\
  {\vrule height 5mm width 0cm}
  $c^{(k)}_{20}$ & 4.708 & 5.541  & 6.593 & 8.686          & 10.43     & 23.15     & 1.215[2]    &
    1.744[2]     \\
  {\vrule height 5mm width 0cm}
  $c^{(k)}_{30}$ & 0.7978 & 0.8784 & 1.454  & 1.800         & 4.686     & 9.673     & 1.440[2]    &
    1.875[2]     \\
  {\vrule height 5mm width 0cm}
  $c^{(k)}_{01}$ & 0.5496 & 0.6676 & 0.3276 & 0.4831        & 0.2152    & 0.5402    & 0.1541   &
    0.3033    \\
  {\vrule height 5mm width 0cm}
  $c^{(k)}_{11}$ & 0.2941 & 0.3170  & 0.2205 & 0.2674        & 0.1939    & 0.1159    & 0.5494   &
    0.5068    \\
  {\vrule height 5mm width 0cm}
  $c^{(k)}_{21}$ & -0.1353 & -0.1392  & -0.1704 & -0.1986      & -0.2374   & -0.4139   & -3.003   &
    -3.612    \\
  {\vrule height 5mm width 0cm}
  $c^{(k)}_{02}$ & 0.02918 & 0.02546  & 0.01543 & 0.01372      & 0.006196  & 0.002691  & 0.007021 &
    0.005165  \\
  {\vrule height 5mm width 0cm}
  $c^{(k)}_{12}$ & 0.01349 & 0.01197  & 0.01133 & 0.01136      & 0.007371  & 0.009352  & 0.03433  &
    0.03460   \\
  {\vrule height 5mm width 0cm}
  $\bar{c}^{(k)}_1$ & 0.152[-3] & 0.133[-3]& 0.432[-4] & 0.317[-4] & 0.165[-4]& 0.156[-4]& 0.343[-6]&
    0.292[-6]\\
  {\vrule height 5mm width 0cm}
  $\bar{c}^{(k)}_2$ & -0.03851 & -0.03411 & -0.02567 & -0.02489    & -0.01329  & -0.01212  & -0.03038 &
    -0.02761  \\
  {\vrule height 5mm width 0cm}
  $\bar{c}^{(k)}_3$ & -5.166 & -5.978     & -9.952 & -12.86        & -33.81    & -75.27    & -1.002[3]   &
    -1.387[3]    \\
  {\vrule height 4mm width 0cm}
  $\delta$, \% & 2.5 & 4.5       & 2.8     & 3.9          & 0.9       & 1.8       & 5.7      &
    4.6
 \end{tabular}
 \end{footnotesize}
 \end{center}
 \end{table}

 \begin{table}[p]
 \begin{center}
 \begin{footnotesize}
 \caption{Coefficients $b^{(k)_i},i=1,\ldots,4,\,k=1,\ldots,4$ in the fit
 of Eqs.~(\ref{eq-fithyd}). Besides the overall r.m.s. deviation $\delta$
 we also give the r.m.s. deviation $\delta^*$, evaluated with Eq.~(\ref{allk})
 by restricting the summation to the points with
 $0.5\le\langle z\rangle\le100$ cm, that are of primary interest.}
 \label{tab-fithyd}
 \begin{tabular}{c|D{.}{.}{7}D{.}{.}{4}D{.}{.}{4}D{.}{.}{6}D{.}{.}{2}D{.}{.}{2}
 @{\hspace{1mm}}|}
% \begin{tabular}{c|rr|rr|rr|rr|}
 {\vrule height 5mm width 0cm}
 &
 \multicolumn{1}{c}{$b^{(k)}_1$} &
 \multicolumn{1}{c}{$b^{(k)}_2$} &
 \multicolumn{1}{c}{$b^{(k)}_3$} &
 \multicolumn{1}{c}{$b^{(k)}_4$} &
 \multicolumn{1}{c}{$\delta$,\%} &
 \multicolumn{1}{c}{$\delta^*$,\%}
  \\
 \hline
 $\langle z\rangle(p,H)\ (k=1) $ & 0.01250 & 1.895 & 0.1379 & 0.009103 &
 2.1 & 2.3 \label{zavzav}\\
 $\sigma_z(p,H)\ (k=2) $ &
 0.0004213 & 3.249 & -0.04348 & 0.2325 & 7.8 & 2.9 \\
 $\langle r\rangle(p,H)\ (k=3) $ & 0.0003259 & 1.792 & 0.2616 &
 0.007800 & 9.8 & 0.9 \\
 $\sigma_r(p,H)\ (k=4) $ &
 0.0003813 & 1.798 & 0.1568 & 0.009332 & 7.0 & 2.3
 \end{tabular}
 \end{footnotesize}
 \end{center}
 \end{table}

 \section{Verification of the results}
 \label{ver}

 Though the results about the propagation of negative muons in
 selected materials presented above were obtained
 exclusively by fitting simulated data
 generated with the widely approved {\sc fluka} code, they
 need further verification by comparison with existing
 experimental and theoretical data.

 Most straightforward is the comparison of
 the mean muon path $\langle z\rangle(p,H)$ of Eq.~(\ref{zavzav})
 with the CSDA range values $z_T$ of
 negative muons in hydrogen\footnote{See Ref.~\cite{groom} for definitions},
 tabulated in \cite{PDG,NPL}.
 For muon energies above 10 MeV (i.e. $p\ge47$ MeV/c)
 we compare with the values of $z_T$  of Ref.~\cite{PDG} that are
 compatible with \cite{NPL}.
 For lower energies the comparison is done with $z_T$ of Ref.~\cite{NPL}
 since the results of \cite{PDG} are estimated ``not dependable'' by the
 authors themselves.
 In Table~\ref{tab-compar} we juxtapose these values of
 $z_T$ with the values of $\langle z\rangle$ obtained using Eq.~(\ref{eq-fithyd}),
 $k=1$. The good agreement between them confirms the validity of
 our results for the free path $\langle z\rangle$ of low-energy muons in gaseous hydrogen.
 To compare with the available results on the muon range in aluminum, steel, gold and
 polystyrene, we take into account that the range value is in fact
 the minimal thickness $d_T$ of a layer of these materials
 for which 100\% of the incident muons with the specified
 initial energy $E_T$ (or, equivalently, initial momentum
 $p_T=c^{-1}\sqrt{E_T(E_T+2m_{\mu}c^2)}$, $m_{\mu}$ being the muon mass\,) are stopped.
 On the other hand, the breakdown momentum $p_0(d)$ was defined in
 Subsection \ref{p0} as the value of the initial momentum for
 which 50\% of the incident muons are stopped in a layer of
 thickness $d$. Since in the vicinity of $p_0$ the dependence of the
 fraction of stopped muons $Q(p,d)$ is very steep (see
 Fig.~\ref{stip2}b), to a good accuracy
 we should expect the following relation to hold:
 \begin{equation}
 p_0(d_T)= p_T. \label{eq-compar}
 \end{equation}
 The agreement between the values of $p_T$ and $p_0(d_T)$ (see Table
 \ref{tab-compar}) confirms the validity of
 the expression of Eq.~(\ref{fitsteep})
 for the breakdown momentum $p_0$.

\begin{figure}[h]
 \begin{center}
 \includegraphics[width=.9\textwidth]{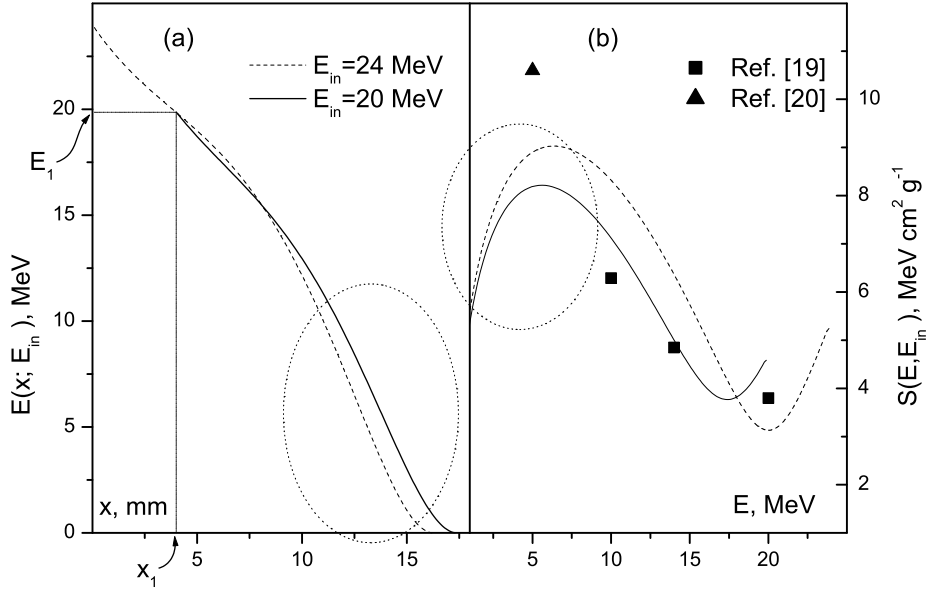}
 \caption{
 (a) Mean energy $E(x,E_{in})$ of monochromatic bunches of muons
 with initial energy 20 and 24 MeV propagating in aluminum, as function
 of the aluminum layer thickness $x$.
 The coordinates $(x_1,E_1)$ of the beginning of the $E(x,E_{in}=20)$
 curve satisfy the relation $E_1=E(x_1,E_{in}=24)$, analogous to
 Eq.~(\ref{relation}).
 For larger $x$, however, the two curves
 deviate significantly.
 (b) Statistical stopping power curves $S(E,E_{in})$ of muons in
 aluminum, evaluated using Eqs.~(\ref{pav}-\ref{sigmath})
 for initial energies $E_{in}=20$ and 24 MeV, juxtaposed with
 the results on the stopping power of Refs.~\cite{PDG,NPL}.
 The discrepancy at small energies (the encircled areas) is
 related to the behavior of $E(x,E_{in})$ near the muon stop point.}
 \label{stopping}
 \end{center}
 \end{figure}

 Comparison with the available data on the
 stopping power $\bar{S}(E)$
 of negative muons in Aluminum from
 Refs.~\cite{PDG,NPL} is not straightforward.
 $\bar{S}(E)$ is defined as
 \begin{equation}
   \bar{S}(E)=-\rho^{-1}\,d\bar{E}(x)/dx,
   \label{stpw-cl}
 \end{equation}
 where $\bar{E}(x)=\bar{E}(x;E_0)$ is the energy of muons of
 initial energy $E_{in}$ at the end of a path of length $x$
 across aluminum with density $\rho$,
 evaluated in the CSD approximation.
 $\bar{E}(x;E_0)$ satisfies the relation
 \begin{equation}
%   \bar{E}(x;\bar{E}(x_1,E_0))=\bar{E}(x+x_1,E_0)
   \bar{E}(x-x_1;E_1)=\bar{E}(x,E_0),
   \text{\ where\ }E_1=\bar{E}(x_1,E_0)
   \label{relation}
 \end{equation}
 for any $E_0$ and $x$, $x_1$
 within the CSDA range.
 What we evaluate instead is the {\em mean energy} $E(x,E_{in})$ of a
 monochromatic bunch of muons with initial energy $E_{in}$ at the
 end of a path of length $x$ or, to be precise, the energy
 $E=\sqrt{\langle p'\rangle^2c^2+m_{\mu}^2c^4} -m_{\mu}c^2$ that
 corresponds to the mean momentum $\langle p'\rangle(x,p)$ of a
 beam with initial momentum $p=c^{-1}\sqrt{E_{in}(E_{in}+2m_{\mu}c^2)}$.
 $E(x,E_{in})$ does not satisfy the relation (\ref{relation}): the evolution
 of the mean energy depends substantially on $E_{in}$, as shown on
 Figure~\ref{stopping}(a).
 The statistical analog of the stopping power of Eq.~(\ref{stpw-cl}),
 defined by $S(E;E_{in})=-\rho^{-1}\,dE(x;E_{in})/dx$,
 also depends on $E_{in}$ and therefore can be compared with $\bar{S}(E)$
 only qualitatively (see Figure~\ref{stopping}(b)). For energies
 of 10 MeV and higher the agreement is reasonable, while for lower energies
 (the encircled area) $S(E;E_{in})$ is smaller than $\bar{S}(E)$ and approaches
 zero as $E\to0$.
 This is due to the fact that in the neighborhood of the breakdown momentum,
 the final momentum distribution $f(p';x,p,{\bf M})$ is
 significantly broadened (see Figure~\ref{al4plots}) in an
 asymmetric way so that most of the muons are stopped beyond the CSDA
 range. This leads in turn to a slower decrease of the
 mean energy
 $E(x,E_{in})$ with $x$ as compared with $\bar{E}(x)$, and lower values of
 $S(E;E_{in})$ in comparison with the stopping power data from Ref.~\cite{NPL}.

 There also are a few direct measurements of the breakdown
 momentum in various materials.
 Ref.~\cite{wilh} reports the experimental value of $6.86$ MeV/c
 for the breakdown momentum in aluminum plate of thickness 0,81
 mg\,cm$^{-2}$. The value 6.12 MeV/c obtained with Eq.~(\ref{fitsteep})
 is in reasonable agreement with experiment. Ref.~\cite{wojc}
 reports the results of measurements of the energy loss of low-energy
 muons in thin layers of carbon and gold. Using their data
 we obtained that muons with initial momentum $p=$1.94 MeV/c
 (the mean exit momentum for the 20 keV muons launched on
 the 3.5 $\mu$ g\,cm$^{-2}$ carbon backing) cross the
 10 $\mu$m thick gold foil with final momentum $p'=1.75$ MV/c,
 while Eq.~(\ref{pav}) gives 1.79 MeV/c, again in good agreement
 with experiment.

 The tentative formula of Eq.~(\ref{fit-step2}) for the breakdown momentum
 was tested for hydrogen and a few more
 solid materials, incl. carbon, nickel, copper and zinc, and
 for low energy muons with momentum up to 75 MeV/c produced results that
 differ from the what is obtained from Refs.~\cite{PDG} by less
 than 5\%.

 Though the angular distribution of muons scattered by various materials
 has also been the subject of experimental investigations (e.g. in
 \cite{muscat}), we did not come across any data that
 could be directly juxtaposed with values obtained with
 Eqs.~(\ref{thav}) and (\ref{sigmath}).

 \begin{table}[h]
 \begin{center}
% \begin{footnotesize}
 \caption{Comparison with data on the CSDA range of low-energy negative muons
 in selected media. $E_T$ and $p_T$
 denote the initial muon energy and momentum (in units MeV and MeV/c,
 respectively). $z_T$ and $\langle z\rangle$ are the values (in
 cm) of the muon path
 in gaseous hydrogen at 1 Atm and 0$^{\circ}$C evaluated
 in the CSDA approximation and using
 Eq.~(\ref{eq-fithyd}), respectively;
 $d_T$ is the muon range (in mm) in the materials of interest,
 and  $p_0(d_T)$ is the breakdown momentum, evaluated using
 Eqs.~(\ref{eq-compar}) and (\ref{fitsteep}).
 The values of $z_T$ and $d_T$ are taken from
 Refs.~\cite{NPL} (for $E_T=1$ MeV) and \cite{PDG} (for higher energies).}
 \label{tab-compar}
 \begin{tabular}{rD{.}{.}{1}|D{.}{.}{1}D{.}{.}{1}|D{.}{.}{1}D{.}{.}{1}|D{.}{.}{2}D{.}{.}{1}|
 D{.}{.}{1}D{.}{.}{1}|D{.}{.}{1}D{.}{.}{1}}
% @{\hspace{1mm}}|}
%
 \\ & &
  \multicolumn{2}{c|}{hydrogen} &
 \multicolumn{2}{c|}{polystyrene} &  \multicolumn{2}{c|}{aluminum} &
 \multicolumn{2}{c|}{steel} &  \multicolumn{2}{c}{gold} \\
% \begin{tabular}{c|rr|rr|rr|rr|}
 \multicolumn{1}{c}{$E_T$} & \multicolumn{1}{c|}{$p_T$} &
 \multicolumn{1}{c}{$z_T$} & \multicolumn{1}{c|}{$\langle z\rangle$} &
 \multicolumn{1}{c}{$d_T$} & \multicolumn{1}{c|}{$p_0(d_T)$} &
 \multicolumn{1}{c}{$d_T$} & \multicolumn{1}{c|}{$p_0(d_T)$} &
 \multicolumn{1}{c}{$d_T$} & \multicolumn{1}{c|}{$p_0(d_T)$} &
 \multicolumn{1}{c}{$d_T$} & \multicolumn{1}{c}{$p_0(d_T)$} \\
 \hline
 {\vrule height 5mm width 0cm}
 1 & 14.6 & 54.4 & 53.3 & & & 0.057 & 14.4 & & &  \\
 10 & 47.0 & 3693. & 3686. & 6.68 & 46.8 & 3.34 & 47.3 & 1.30 & 47.6 &
 0.48 & 48.7 \\
 14 & 56.2 & 6729. & 6757. & 12.2 & 55.6 & 6.08 & 56.4 & 2.36 & 56.7 &
 1.37 & 57.9 \\
 20 & 68.0 & 12627. & 12774. & 22.9 & 66.4 & 11.3 & 67.5 & 4.36 & 68.1 &
 2.49 & 69.3
 \end{tabular}
% \end{footnotesize}
 \end{center}
 \end{table}

 \section{Discussion of the results}

 We start by stressing once again that the
 results presented here are not aimed at substituting any
 full scale Monte Carlo simulations but only at helping the
 early stage design of
 the set-up for experiments where stopping and capture of
 low energy muons is studied.
 Knowing the details of the different types of processes
 has proven to be useful in
 restricting the range of the various parameters that are subject to
 optimization, and significantly enhances the efficiency
 of the full scale simulations. In what follows we exemplify the
 usefulness of our approach.

\begin{figure}[ht]
 \begin{center}
 \includegraphics[width=.8\textwidth]{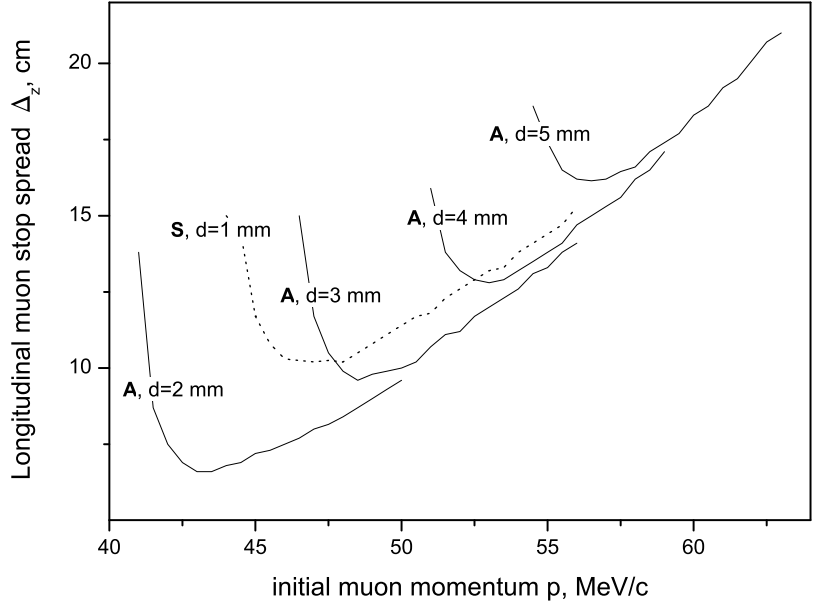}
 \caption{FWHH longitudinal size of the muon stopping area
 in pure hydrogen at 40 Atm and 300 K for a monochromatic collinear
 muon beam that has crossed aluminum plates of thickness 2(1)5
 mm or a steel plate of thickness 1 mm.}
 \label{spread}
 \end{center}
 \end{figure}

 1. Consider the distribution of the muon stopping
 points along the axis $z$. Figure \ref{scat}(b) shows the shape
 of the distribution density $S(z;p,H)$ under the assumption that all muons
 enter the gas target with the same momentum $p$. In fact, after
 crossing the wall of the gas container the incident muon beam is
 no longer monochromatic and collinear; the distribution density
 in this case becomes
 \begin{equation}
   \tilde{S}(z;p,H)\equiv\tilde{S}(z;p,H,d,{\bf M})=
   \int f(p';p,d,{\bf M})\,S(z;p',H)\,dp',
   \label{tilde}
 \end{equation}
 where $f(p';p,d,{\bf M})$ is the final momentum distribution
 density for muons, launched with initial momentum $p$\, against a
 layer of {\bf M} with thickness $d$ mm. We approximated it
 with the normal distribution density
 \begin{equation}
  f(p';p,d,{\bf M})\simeq N(\langle p'\rangle(p,d,{\bf M}),
  \sigma_{p'}(p,d,{\bf M}))
 \end{equation}
 and similar for $s(z;p,H)$, performed numerically the integration
 in (\ref{tilde}) and evaluated the FWHH longitudinal spread of the
 muon stopping area $\Delta_z(p,d,H,{\bf M})$. Figure \ref{spread}
 displays the dependence of $\Delta_z(p,d,H,{\bf M})$ on the initial
 momentum $p$ for aluminum plates with thickness 2(1)5 mm and a
  1 mm steel plate. The curves have distinct minima for which
 the muons stopping area is most compact as needed to satisfy
 Cond.~2, discussed in the Introduction:
 appropriate positioning the detectors that signal the formation of muonic atoms
 by registering the characteristic X-rays
 will maximize their efficiency. The full
 scale MC search of the optimal initial momentum and
 detector positions may thus be
 restricted to a narrower range.

\begin{figure}[ht]
 \begin{center}
 \includegraphics[width=.8\textwidth]{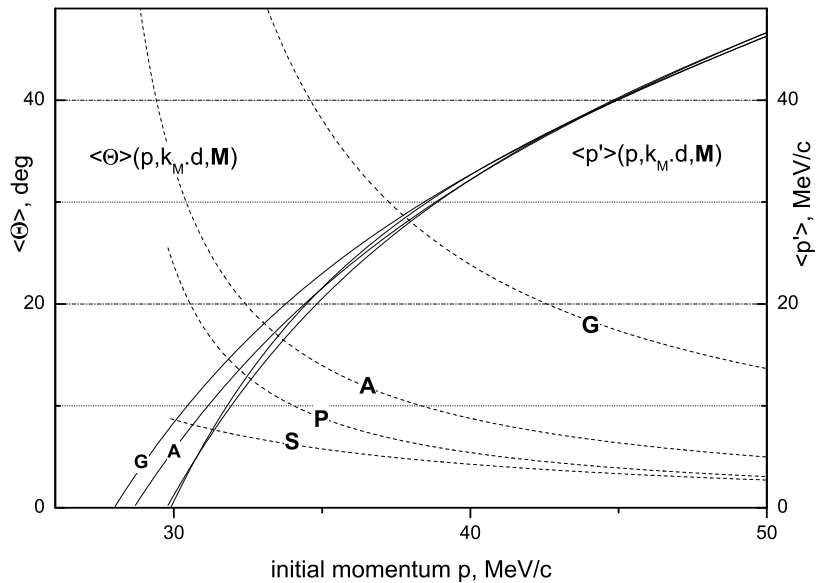}
 \caption{The plots of the mean final momentum
 $\langle p' \rangle$ of muon beams that have crossed a 1 mm
 layer of aluminum {\bf A} or a layer of other materials
 {\bf P, S, G} with thickness,
 rescaled according to (\ref{eq-scaled}) (the solid lines)
 approximately coincide for $p\ge38$. On the contrary, the mean
 deviation angles $\langle\Theta\rangle$ (the dashed lines)
 are very different and not
 directly correlated with the material density.
 }
 \label{scaled}
 \end{center}
 \end{figure}

 2. The expressions $\langle p'\rangle(p,d,{\bf M})$ for the mean final
 momentum approximately satisfy the following scaling relations:
 \begin{equation}
 \langle p'\rangle(p,k_{\bf M}d,{\bf M})\approx\langle p'\rangle(p,k_{\bf M'}d,{\bf M'}),
   \label{eq-scaled}
 \end{equation}
 valid for values of $p$ above the breakdown momentum $p_0$ by
 5-10 MeV/c and higher (see Figure
 \ref{scaled}). We empirically determined the following values of the
 material-dependent coefficients $k_{\bf M}$: $k_{\bf A}=1$,
 $k_{\bf P}=2.031$, $k_{\bf S}=0.3851$, and $k_{\bf G}=0.2138$.
 The mean values of the angle of deviation $\langle
 \Theta\rangle(p,d,{\bf M})$, however, are not scaled even
 approximately. Knowing the angular profile of the muon beam after
 crossing the entrance window of the gas target is of importance
 for reducing the losses of muons in the side walls. As long as the
 angle of deviation in hydrogen gas -- of the order of $2^{\circ}$ --
 is much smaller and can be neglected compared to the deviation
 angle in solid material layers, the preliminary estimate of these losses
 can be done using Eqs.~(\ref{thav},\ref{sigmath}).

 In a concluding remark we note that, although the detailed study
 of the propagation of slow muons using the {\sc fluka} code was
 restricted here to a few media of specific interest,
 the apparently wider validity of Eq.~(\ref{fit-step2}) makes us believe
 that the same approach can be efficiently applied to a much
 broader range of solids and gases, with minimal modifications
 (if any) in the explicit form of the approximating expressions
 (\ref{pav}-\ref{sigmath}) and (\ref{eq-fithyd}).

\acknowledgments

The authors are grateful to Dr. G.~Battistoni for his help on
 {\sc fluka} topics. A.A. and D.B. acknowledge the partial support from
the bilateral agreement of the Bulgarian Academy of sciences and
the Polish Academy of Sciences.

\end{document}